\documentclass[twocolumn,hyperpdf,amsmath,amssymb,aps,prd,10pt,superscriptaddress,nofootinbib,noeprint,preprintnumbers,floatfix]{revtex4-1}
\pdfoutput=1

\usepackage{graphicx, color}

\usepackage[letterspace=-10]{microtype} 

\usepackage{bm, amsmath, amsfonts}
\usepackage{multirow, tabularx, dcolumn}
\usepackage{mathtools} 
\usepackage{booktabs}
\usepackage{placeins}

\usepackage[utf8]{inputenc} 
\usepackage{hyperref}

\usepackage{xcolor}
\definecolor{jlab_red}{RGB}{192,39,45}
\definecolor{jlab_orange}{RGB}{249,102,0}
\definecolor{jlab_blue}{RGB}{47,122,121}
\definecolor{jlab_green}{RGB}{65,125,10}

\definecolor{dstarpi_s}{RGB}{191,39,45}
\definecolor{dstarpi_d}{RGB}{65,125,10}
\definecolor{dstarpi_mix}{RGB}{248,102,0}
\definecolor{dpi_d}{RGB}{51,92,129}

\newcommand{\cm}{\ensuremath{\mathsf{cm}}}
\newcommand{\tm}{\ensuremath{\text{-}}}

\newcommand{\osz}{\ensuremath{\:\!^1\!S_0}}
\newcommand{\tso}{\ensuremath{\:\!^3\!S_1}}
\newcommand{\tpz}{\ensuremath{\:\!^3\!P_0}}
\newcommand{\opo}{\ensuremath{\:\!^1\!P_1}}
\newcommand{\tpo}{\ensuremath{\:\!^3\!P_1}}
\newcommand{\tpt}{\ensuremath{\:\!^3\!P_2}}
\newcommand{\tdo}{\ensuremath{\:\!^3\!D_1}}
\newcommand{\odt}{\ensuremath{\:\!^1\!D_2}}
\newcommand{\tdt}{\ensuremath{\:\!^3\!D_2}}

\hypersetup{%
pdftitle = {Axial-vector D_1 hadrons in Dstar-pi scattering from QCD},
pdfsubject = {QCD},
pdfkeywords = {QCD, Hadron, Charm, Physics, Lattice, Meson, Scattering},
pdfauthor = {Hadron Spectrum Collaboration},
colorlinks = {true},
filecolor = {black},
linkcolor = {jlab_blue},
menucolor = {black},
citecolor = {jlab_green},
urlcolor = {jlab_green},
}{}

\begin{document}


\title{Axial-vector $D_1$ hadrons in $D^\ast\pi$ scattering from QCD}

\author{Nicolas Lang}
\email{nicolas.lang@maths.tcd.ie}
\affiliation{School of Mathematics, Trinity College, Dublin~2, Ireland}
\author{David~J.~Wilson}\email{d.j.wilson@damtp.cam.ac.uk}
\affiliation{DAMTP, University of Cambridge, Centre for Mathematical Sciences, Wilberforce Road, Cambridge, CB3 0WA, UK}

\collaboration{for the Hadron Spectrum Collaboration}
\date{\today}

\begin{abstract}
We present $I=1/2$ $D^\ast\pi$ scattering amplitudes from lattice QCD and determine two low-lying $J^P=1^+$ axial-vector $D_1$ states and a $J^P=2^+$ tensor $D_2^\ast$. Computing finite-volume spectra at a light-quark mass corresponding to $m_\pi=391$ MeV, for the first time, we are able to constrain coupled $J^P=1^+$ $D^\ast\pi$ amplitudes with $^{2S+1}\ell_J\,=\,^3S_1$ and $^3\!D_1$ as well as coupled $J^P=2^+$ $D\pi\{^1\!D_2\}$ and $D^\ast\pi \{^3\!D_2\}$ amplitudes via L\"uscher's quantization condition. Analyzing the scattering amplitudes for poles we find a near-threshold bound state, producing a broad feature in $D^\ast\pi\{^3\!S_1\}$. A narrow bump occurs in $D^\ast\pi\{^3\!D_1\}$ due to a $D_1$ resonance. A single resonance is found in $J^P=2^+$ coupled to $D\pi$ and $D^\ast\pi$.  A relatively low mass and large coupling is found for the lightest $D_1$, suggestive of a state that will evolve into a broad resonance as the light quark mass is reduced. An earlier calculation of the scalar $D_0^\ast$ using the same light-quark mass enables comparisons to the heavy-quark limit.
\end{abstract}

\maketitle

\noindent
\emph{Introduction} --- 
Since their discovery, the lightest scalar $D$-meson excitations have raised questions about their mass ordering, widths and composition. Heavy-quark spin symmetry suggests the same questions apply to the axial vectors. At the same time, the $D$-meson resonances are exemplary for several charmed states~\cite{Belle:2003nnu,LHCb:2020bwg,LHCb:2020bls,LHCb:2020pxc}, some manifestly exotic, arising in coupled-channel systems and close to thresholds. They may therefore serve as a place to obtain a more general understanding of QCD dynamics among charmed hadrons.

In experiment, there are four low-lying positive-parity $D$-mesons~\cite{ParticleDataGroup:2020ssz}: a scalar $D_0^\ast(2300)$, two axial-vectors $D_1(2430)$ and $D_1(2420)$, and a tensor $D_2^\ast(2460)$. The scalar and $D_1(2430)$ are very broad and are thought to couple strongly to their respective $D\pi$ and $D^\ast\pi$ decay modes. The $D_1(2420)$ and the tensor $D_2^\ast$ are relatively narrow, the $D_2^\ast$ decaying into both $D\pi$ and $D^\ast\pi$.

A range of theoretical approaches have been applied to understand these states. Quark potential models~\cite{Godfrey:1985xj} provide a useful qualitative starting point leading to four states arising from the $\ell=1$ singlet $^1\!P_1$ and triplet $^3\!P_J$ combinations. Charge-conjugation is not a good symmetry so $\opo$ and $\tpo$ mix. This produces one $J^P=0^+$, two $1^+$ and a $2^+$ state at similar masses~\cite{Godfrey:1985xj}. Approaches considering the coupling to decay channels~\cite{Godfrey:1986wj,Bardeen:1993ae,vanBeveren:2003jv,Kolomeitsev:2003ac,Hofmann:2003je,Burns:2014zfa} are necessary when strong $S$-wave decay modes are present. Recent studies have shown that the scalar $D_0^\ast$ pole may be far below its currently reported value~\cite{Bardeen:1993ae,vanBeveren:2003kd,Bardeen:2003kt,Albaladejo:2016lbb,Du:2020pui,Gayer:2021xzv}, and the same could be true of the broad $D_1$.

One particularly useful theoretical perspective is obtained by considering the behaviour when the charm quark becomes infinitely heavy with respect to both the light quarks and the scale of QCD interactions~\cite{DeRujula:1976ugc,Rosner:1985dx,Godfrey:1986wj,Isgur:1991wq,Lu:1991px,Bardeen:2003kt,Godfrey:2005ww,Colangelo:2012xi}. In this limit the spin of the heavy quark is conserved, and the $D$-meson states can be characterized by the vector sum of the orbital angular momentum and the light-quark spin. For the quark-model $\ell=1$ states two doublets arise. One doublet contains the $D_0^\ast$ and one of the $D_1$ mesons, and in the infinitely heavy-quark limit they decay exclusively via $S$-wave interactions. The other contains a $D_1$ and the $D_2^\ast$ decaying entirely by $D$-wave interactions. 

Recent advances have enabled computations of the properties of hadron resonances using \emph{lattice} QCD. Evidence for highly excited $D$-mesons has been obtained, including patterns of states beyond the quark model with apparent gluonic content~\cite{Liu:2012ze,Moir:2013ub,Cheung:2016bym}. These methods are able to determine the scattering amplitudes and their spectroscopic content to compare with experiment. Relevant for the axial-vector $D_1$ mesons, the scattering of hadrons with spin in coupled $\tso$--$\tdo$ amplitudes was first studied in a weakly-interacting system~\cite{Woss:2018irj} and later in the context of the $b_1$ resonance~\cite{Woss:2019hse}. 
The scalar charmed resonances have been investigated in both the charm-light~\cite{Mohler:2012na,Moir:2016srx,Gayer:2021xzv} and charm-strange~\cite{Liu:2012zya,Mohler:2013rwa,Lang:2014yfa,Bali:2017pdv,Alexandrou:2019tmk,Cheung:2020mql} flavors. A large coupling to the strong-decay channels was found in both cases. The axial-vectors were also studied in Refs.~\cite{Mohler:2012na,Lang:2014yfa,Bali:2017pdv}.

In this article, we determine the $D_1$ masses and couplings to $I=1/2$ $D^\ast \pi$ scattering from QCD. We also compute the $D_2^\ast$ and its couplings to both $D^\ast\pi$ and $D\pi$ decay modes. With the $D_0^\ast$ determined from the same lattices~\cite{Moir:2016srx,Gayer:2021xzv}, we compare these states to experiment and probe the predictions from the heavy-quark limit.


\begin{figure*}[!thb]
\includegraphics[width=0.93\textwidth]{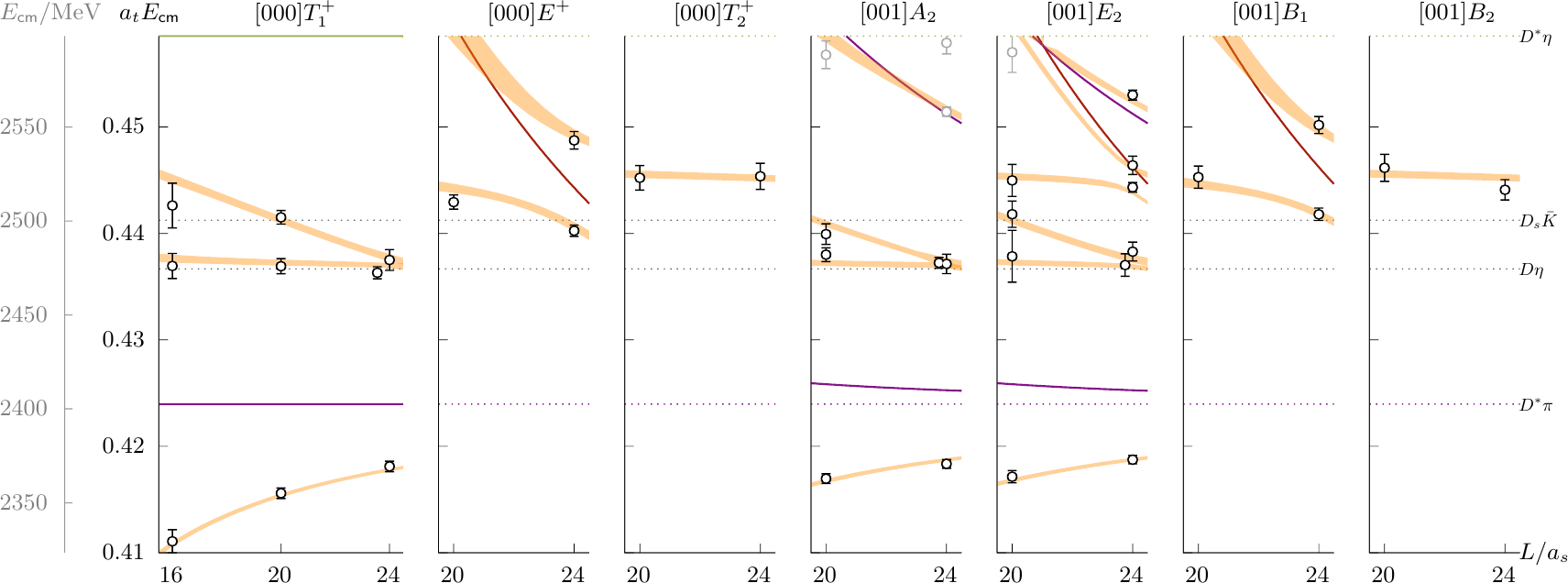}
\caption{The black and grey points show the finite volume spectra obtained from irreps up to $|\vec{P}| = 1$ that contain $J^P \in \{1^+, 2^+\}$. Only black points are used in the fit. The thin solid curves show the spectrum anticipated in the absence of interactions between $D^\ast\pi$ (purple), $D\pi$ (red) and $D^\ast\eta$ (green). Dotted lines represent kinematic meson-meson thresholds. The $D \pi$ threshold lies below the displayed energy range. The light yellow bands correspond to solutions of the determinant condition Eq.~\ref{eq_det} using the described parameterization. Additional irreps are shown in the supplemental~\ref{supp:spectra}.}

\label{fig_spec}
\end{figure*}


\emph{Computing finite-volume spectra} --- Lattice QCD is a numerical approach through which first-principles predictions of strongly-coupled QCD can be obtained by Monte Carlo sampling the QCD path integral. Working in a discretized finite euclidean volume $L^3 \times T$ with spatial and temporal lattice spacings $a_s$ and $a_t$, correlation functions can be computed that determine the QCD spectrum in that volume.

The rotational symmetry of the finite cubic spatial boundary differs from an infinite volume. At rest, we compute spectra within irreducible representations (irreps) of the finite cubic group, rather than the continuous orthogonal group. This results in a linear combination of multiple partial-waves of definite $J^P$ within each irrep. Amplitudes grow at threshold like $k^{2\ell}$ which commonly suppresses all but the lowest few partial waves at low energies. We also consider non-zero overall momentum $\vec{P}=\frac{2\pi}{L}(i,j,k)=[ijk]$, where the symmetry is further reduced; notably, partial waves with both parities are present in the corresponding irreps. The partial waves contributing to the irreps used in this calculation are described in Ref.~\cite{Woss:2018irj}. Irreps are labeled by $[ijk]\Lambda^{(P)}$.

We use lattices with $2+1$ dynamical flavors of quark, where the strange quark is approximately physical and the light quark produces a pion with $m_\pi\approx 391$~MeV. The ratio of the spatial and temporal lattice spacings on this anisotropic lattice is $a_s/a_t\approx 3.5$~\cite{Edwards:2008ja,Lin:2008pr}. We use three volumes with $(L/a_s)^3\times T/a_t=\{16^3,20^3,24^3\}\times 128$. The scale is estimated using the $\Omega$ baryon leading to $a_t^{-1}=5667$~MeV~\cite{Edwards:2012fx}. This corresponds to $a_s\approx0.12\: \mathrm{fm}$, resulting in physical volumes between $(2\:\mathrm{fm})^3$ and $(3\:\mathrm{fm})^3$.  The distillation approach is used, which both enhances signals from the required low energy modes, and allows all of the Wick contractions specified by QCD to be computed efficiently~\cite{Peardon:2009gh}. The charm-quark uses the same action as the light and strange quarks~\cite{Liu:2012ze}. Crucial to this study, the vector $D^\ast$ is stable at this light-quark mass~\cite{Moir:2013ub}.

The variational method is used to extract spectra from correlation functions~\cite{Michael:1985ne,Luscher:1990ck,Blossier:2009kd}. Special care is needed in choosing operators with good overlap onto all the states present within the investigated energy range. In this work, we use a large basis of approximately-local $q\bar{q}$- and meson-meson-like operators. The latter are constructed from pairs of mesons obtained variationally from large bases of $q\bar{q}$-like operators to reduce excited state contributions~\cite{Dudek:2010wm,Thomas:2011rh,Dudek:2012gj}.
The correlation functions form a matrix that is diagonalized using a generalized eigenvalue approach. The time dependence of the eigenvalues yields the finite volume spectrum $\{E_\mathfrak{n}\}$. These spectra expose the underlying scattering amplitude through L\"uscher's finite volume quantization condition, and extensions thereof~\cite{Luscher:1990ux, Rummukainen:1995vs,Bedaque:2004kc,Kim:2005gf,Fu:2011xz,Leskovec:2012gb,Gockeler:2012yj,He:2005ey,Bernard:2010fp, Doring:2012eu,Hansen:2012tf,Briceno:2012yi,Guo:2012hv,Briceno:2014oea} -- these methods are reviewed in Ref.~\cite{Briceno:2017max}.


In Fig.~\ref{fig_spec} we present the finite volume spectra computed in irreps with $\vec{P}=[000]$ and $[001]$ with contributions from $J^P=1^+$ and $2^+$. In $[000]T_1^+$ and $[001]A_2$, we observe 3 energy levels below $a_tE=0.45$ ($\approx 2550$ MeV) where only a single energy level would have been expected based on the non-interacting spectrum. This indicates non-trivial interactions in $J^P=1^+$. The $[000]E^+$ and $[000]T_2^+$ irreps also contain an extra energy level, suggestive of significant $J^P=2^+$ interactions. Further irreps and a list of the operators used are given in supplemental \ref{supp:spectra} and \ref{supp:ops}.


\emph{Scattering amplitude determinations} --- 
The extension of L\"uscher's determinant condition for the scattering of hadrons with arbitrary spin can be written~\cite{Briceno:2014oea}, 
\begin{align}
\det\Bigl[\bm{1}+i \bm{\rho}(E)\cdot\bm{t}(E)\cdot\bigl(\bm{1}+i\bm{\mathcal{M}}(E,L)\bigr)\Bigr]=0,
\label{eq_det}
\end{align}
where $E$ is the $\cm$ energy, $L$ is the spatial extent, $\bm t$ is the infinite volume scattering $t$-matrix and $\bm{\mathcal{M}}$ is a matrix of known functions of energy, which is dense in partial waves and dependent on the irrep. $\bm{\rho}$ is a diagonal matrix of phase-space factors. Since $\bm t$ has multiple unknowns at each value of energy, it is necessary to introduce a parameterization. The $J^P=1^+$ wave can be parameterized using a symmetric $2\times 2$ $K$-matrix, indexed by the $\tso$ and $\tdo$ channel labels. A threshold factor is included to promote the natural threshold behaviour of each $t$-matrix element,
\begin{align}
t^{-1}_{ij} = (2k_i)^{-\ell_i}\: K_{ij}^{-1}\:(2k_j)^{-\ell_j} + I_{ij}\;,
\label{eq_tmat}
\end{align}
where $k_i$ are the $\cm$ momenta, $K_{ij}$ are the $K$-matrix elements and $I_{ij}$ is a diagonal matrix. To respect unitarity, $\mathrm{Im} I_{ij}=-\rho_i\delta_{ij}$, where $\rho_i=2k_i/E$ are the phase space factors. The real part is either set to zero or a Chew-Mandelstam phase space is used~\cite{Wilson:2014cna}, where a logarithm is generated from the known imaginary part.

\begin{figure}[!t]
\includegraphics[width=0.9\columnwidth]{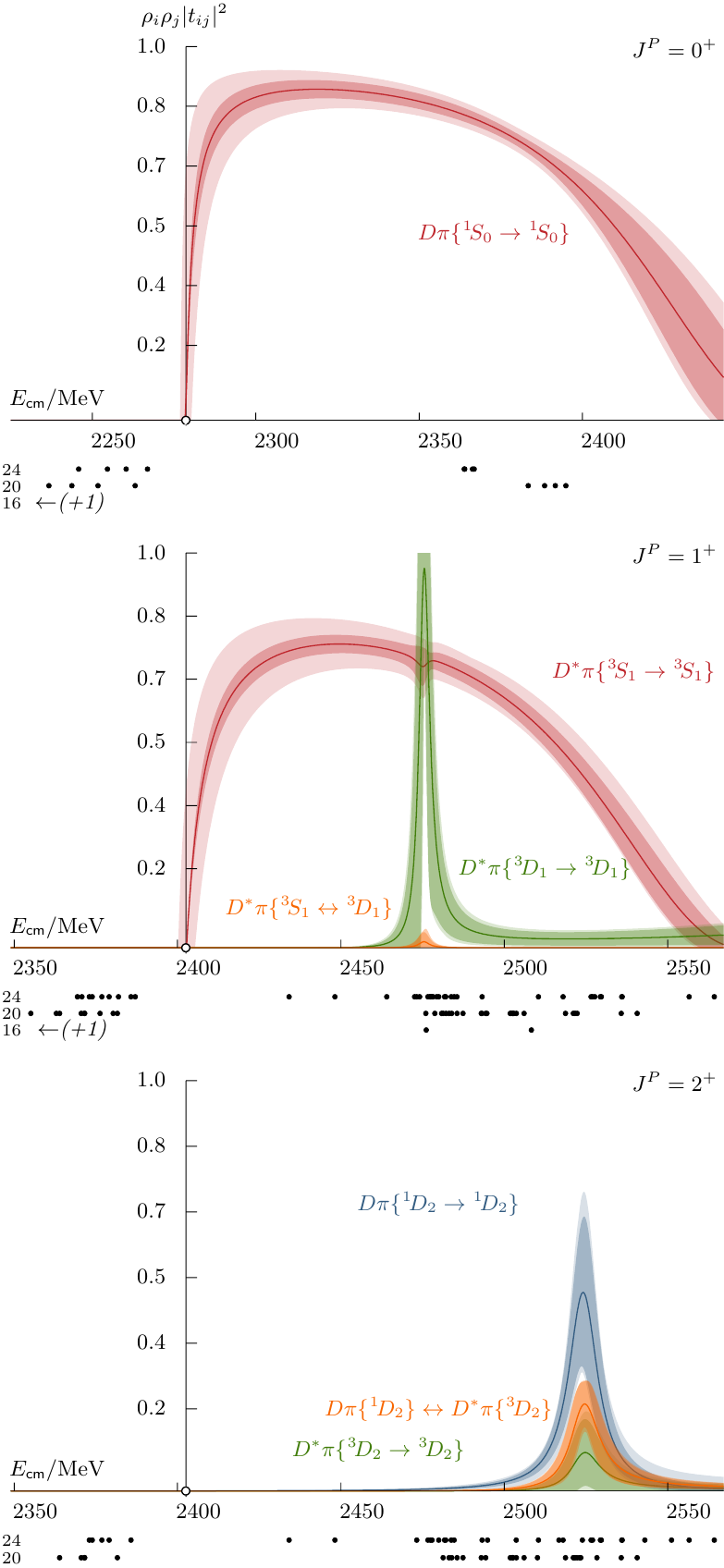}
\caption{
Scattering amplitudes for $J^P \in \{0^+$, $1^+$, $2^+\}$. The outer uncertainty bands are obtained by varying the hadron masses and anisotropy entering eq. \ref{eq_det} within their uncertainties. The energy levels from irreps with a contribution from the respective $J^P$ are shown in black below the horizontal axis. The $0^+$ amplitude determination is from Refs.~\cite{Moir:2016srx,Gayer:2021xzv}.
	 }
\label{fig_amps}
\end{figure}

One useful form of $\bm{K}$ is
\begin{align}
K_{ij}=\sum_{p=1}^{2}\:\frac{g_{p,i}g_{p,j}}{m_p^2-s}+\gamma_{ij}\;.
\label{eq_Kmat}
\end{align}
The $K$-matrix pole mass $m_p$ and couplings $g_{p,i}$ are free parameters that can efficiently produce resonances in a $t$-matrix. The $\gamma_{ij}$ form a real symmetric matrix of constants. The free parameters are determined in a $\chi^2$ minimization as defined in Eq.~8 of Ref.~\cite{Wilson:2014cna}, to find an amplitude that best describes the finite volume spectra obtained in the lattice calculation. We choose to search for the solutions of Eq.~\ref{eq_det} using the eigenvalue decomposition method outlined in Ref.~\cite{Woss:2020cmp}, which is ideally suited to problems with multiple channels and partial waves. 

We determine the $J^P=(1,2)^+$ and $(0,1,2)^-$ partial waves (everything up to $\ell=2$) simultaneously from the irreps in Fig.~\ref{fig_spec}, plus $[011]A_2,B_1,B_2$, $[111]A_2,E_2$ and $[002]A_2$, resulting in 94 energy levels ($J^P = 0^+$ does not contribute to any of these irreps). Each wave is parameterized using a version of Eq.~\ref{eq_Kmat} with various parameters fixed to zero. For example, for the $1^+$ wave we use the poles and the $\gamma_{\tso,\tso}$ element, and fix the other $\gamma_{ij}$ to zero. We use a single pole for the $2^+$ amplitudes, and only constants $\gamma$ for the $J^-$ waves. Further details of the parameterization and the parameter values resulting from the $\chi^2$-minimization are given in the supplemental~\ref{sec:ref_param}. A $\chi^2/N_{\text{dof}} = \tfrac{95.0}{94 - 15} =  1.20$ is obtained. The spectra from using these amplitudes in Eq.~\ref{eq_det} are shown as orange curves in Fig.~\ref{fig_spec}. The amplitudes are plotted in Fig.~\ref{fig_amps}.

The parameterization given in Eq.~\ref{eq_Kmat} is one of many reasonable choices. In order to reduce possible bias from a specific choice, we vary the form. We obtain 21 different parameterizations that describe the spectra, summarized in supplemental~\ref{supp:par_var}. These are used when computing pole positions, couplings and their uncertainties. 

\emph{Poles \& Interpretation} --- 
Scattering amplitude poles describe the spectroscopic content consisting of both unstable resonances and stable bound states. The amplitude is analytically continued to complex-$s$. There is a square-root branch cut beginning at each threshold leading to sheets that can be labelled by the sign of the imaginary part of the momentum in each channel $i$, $\mathrm{Im}k_i$. Close to a pole, the $t$-matrix is dominated by a term $t_{ij} \sim \frac{c_i c_j}{s_0-s}$, where $s_0$ is the pole position and $c_i$ are the channel couplings. 

\begin{figure}[!t]
\includegraphics[width=0.93\columnwidth]{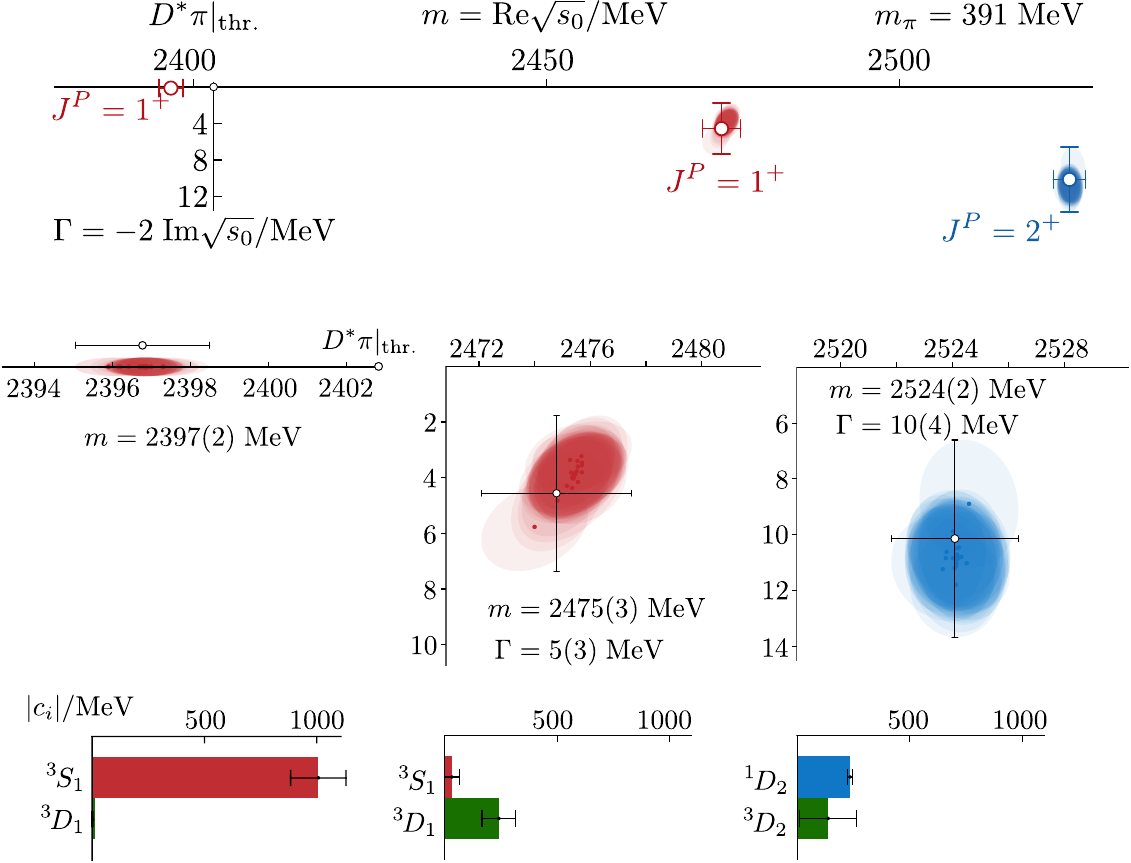}
\caption{The $t$-matrix poles and couplings determined in this study. The top and middle shows pole positions; the bottom shows couplings. The colored error ellipses are the results from individual parameterizations and are used to determine the quoted envelope over all parameterizations shown in black. }
\label{fig_poles}
\end{figure}

In the $J^P=1^+$ amplitudes, we find a bound-state close to threshold, strongly coupled to the $\tso$ amplitude, that produces a broad enhancement over the energy region where the amplitudes are constrained.  The $\tdo$ amplitude has a coupling to this pole consistent with zero, as shown in the lower left panel of Fig.~\ref{fig_poles}. Due to the proximity of the bound state pole to threshold, the $k_i^{\ell_i}$ factor severely dampens any $\tdo$ coupling. The narrow peak in the $\tdo$ amplitude is produced by a resonance pole dominantly coupled to the $\tdo$ amplitude, with only a small $\tso$ coupling. We find a small but non-zero width in all but two of the parameterizations. Both are rejected due to a larger $\chi^2$ than the rest of the amplitudes; the more subtle case is described further in supplemental~\ref{sec:discarded_param}.

An additional pole is found at the upper edge of the fitting range close to where the $\tso$ amplitude touches zero, as seen in Fig.~\ref{fig_amps}. When a zero occurs on the physical sheet, it is usually accompanied by a pole on an unphysical sheet at a similar energy. Since this pole is so close to both the upper limit of the energy range and the $D^\ast\eta$ an $D_s^\ast\bar{K}$ channel openings, higher energy levels are required to be certain of its presence. In one of the rejected amplitudes, this pole does not arise. A similar feature occurs in $D\pi\{\osz\}$ elastic scattering that either disappears or moves to higher energies once the coupled-channel $D\eta$ and $D_s\bar{K}$ amplitudes are introduced~\cite{Moir:2016srx}. We thus exclude this pole from the remaining discussion.

One pole is found in $J^P=2^+$, coupled to both $D\pi\{\odt\}$ and $D^\ast\pi\{\tdt\}$. No other nearby poles are found.

In summary, poles are found at
\begin{center}
\noindent
{
\begin{tabular}{llll}
$\sqrt{s_{D_0^\ast}}$    &=  $2276(2) \,\mathrm{MeV}\quad(+)$~(from Refs.~\cite{Moir:2016srx,Gayer:2021xzv})\\
$\sqrt{s_{D_1}}$         &=  $2397(2) \,\mathrm{MeV}\quad(+)$\\
$\sqrt{s_{D_1^\prime}}$  &= $\left(2475(3) -\tfrac{i}{2} 5(3) \right)\,\mathrm{MeV} \;\;\quad(-)$\\
$\sqrt{s_{D_2^\ast}}$    &= $\left(2524(2) -\tfrac{i}{2} 10(4) \right)\,\mathrm{MeV} \quad(-,-)\,.$
\end{tabular}
}
\end{center}
\noindent
The signs in curved brackets indicate the sign of $\mathrm{Im}k_i$, and thus the sheet where the pole is located. The quoted values correspond to the envelope of uncertainties of all accepted parameterizations as well as mass and anisotropy variations.
The couplings are (in MeV):
\begin{center}
\noindent
{
\begin{tabular}{llll}
$D_0^\ast$   & $\;|c_{D\pi\{\osz\}}|$       $= $ 760(164) & (from Refs.~\cite{Moir:2016srx,Gayer:2021xzv})\\
$D_1$        & $\;|c_{D^\ast\pi\{\tso\}}|$  $= $ 1007(123) & $|c_{D^\ast\pi\{\tdo\}}| = 2(3)$ \\
$D_1^\prime$ & $\;|c_{D^\ast\pi\{\tso\}}|$  $= $ 32(33) & $|c_{D^\ast\pi\{\tdo\}}| =  239(77)$ \\
$D_2^\ast$   & $\;|c_{D\pi\{\odt\}}|$       $= $ 234(11) & $|c_{D^\ast\pi\{\tdt\}}| =  137(126)$\,. \\
\end{tabular}
\vspace{0.1cm}
}
\end{center}

Comparing to experiment, the larger-than-physical light-quark mass must be accounted for. $D$-meson masses larger than those found in experiment are expected. The $D_2^\ast$ is found some 80 MeV above the experimental state and the narrow $D_1^\prime$ is 52 MeV above experiment. The bound-state pole that produces the broad feature in $S$-wave however is 15~MeV \emph{below} the $D_1(2430)$ found at 2412(9) MeV~\cite{LHCb:2019juy,ParticleDataGroup:2020ssz}. Considering the similarity of $D\pi\{\osz\}$ and $D^\ast\pi\{\tso\}$ along with the light-quark mass dependence of the $D_0^\ast$ determined  in Ref.~\cite{Gayer:2021xzv}, the broad $D_1(2430)$ could in reality be produced by a pole at a much lower mass. 
Another difference that arises as the light-quark mass is reduced, is that $D\pi\pi$ opens and may introduce more mixing between channels. Recent developments of three-body formalism will be essential in understanding these processes~\cite{Hansen:2015zga,Briceno:2018aml,Hansen:2019nir,Blanton:2020jnm,Blanton:2020gmf,Blanton:2021mih,Blanton:2021eyf,Mai:2017bge,Polejaeva:2012ut,Hammer:2017uqm,Hammer:2017kms}.

Crudely extrapolating the couplings to lighter pion masses requires a kinematic factor, leading to $|c^\mathrm{ex.}|=|c||k^\mathrm{ex.}(m_r^\mathrm{ex.})/k(m_r)|^\ell$~\cite{Wilson:2015dqa,Woss:2020ayi}. The $D_1^\prime$ couplings, fixing $m_{D_1^\prime}^{\mathrm{ex.}}=2422$~MeV~\cite{ParticleDataGroup:2020ssz}, produce a consistent width to that observed experimentally, $\Gamma_{D_1^\prime}=48\pm 30$~MeV. The $D_2^\ast$ couplings, fixing $m_{D_2^\ast}^{\mathrm{ex.}}=2461$~MeV~\cite{ParticleDataGroup:2020ssz}, correspond to $\Gamma_{D_2^\ast}=24\pm 13$~MeV, a little narrower than that observed in experiment. The large coupling found for the bound $D_1$ suggests this will evolve into a broad resonance as the light quark mass is reduced.

$D^\ast K$ in $I=0$ is related to $D^\ast\pi$ in $I=1/2$ by SU(3) flavor symmetry~\cite{Albaladejo:2016lbb,Cheung:2020mql} and has similarities. The experimental axial-vector $D_{s1}$ hadron masses follow a similar pattern to that observed here, with one bound state and one narrow resonance. The bound-state is significantly more-bound, and the resonance is narrower. Evidence of a bound $D_{s1}$ was also found in other lattice studies~\cite{Lang:2014yfa,Bali:2017pdv}.

 
\emph{Heavy-quark limit comparisons} --- 
The heavy quark limit (as described in Refs.~\cite{DeRujula:1976ugc,Rosner:1985dx,Godfrey:1986wj,Isgur:1991wq,Lu:1991px,Bardeen:2003kt,Godfrey:2005ww}) captures many of the features observed in these states. Notably, the prediction of decoupling between the two $1^+$ states is upheld in our results. We could not rule out a $\tso$ component in the narrow $D_1^\prime$ resonance; many of the parameterizations favor a small but non-zero coupling. The $D\pi\{\osz\}$ and $D^\ast\pi\{\tso\}$ amplitudes are remarkably similar in terms of both singularities and energy dependence, producing near-threshold bound states strongly coupled to the relevant nearby channel. Binding energies are found to be $2(1)$~MeV and $6(2)$~MeV respectively. Heavy-quark symmetry can also be used to relate the couplings of the narrow $D_1^\prime$ and $D_2^\ast$. The results found here are not inconsistent with expectations from the heavy-quark limit; a proper test needs a more precise determination of the couplings, in particular $c_{D^\ast\pi\{\tdt\}}$.


\emph{Summary} --- The dynamically-coupled $^{2S+1}\ell_J=\tso$ and $\tdo$ $D^\ast\pi$ scattering amplitudes in $I=1/2$ have been computed from QCD for the first time. Working at $m_\pi=391$~MeV, the $D^\ast\pi\{\tso\}$ amplitude is dominated by a pole just below threshold, whose influence extends over a broad energy region. This pole is found below the experimental state despite the larger pion mass, analogous to the scalar sector. There is a narrow resonance coupled dominantly to the $D^\ast\pi\{\tdo\}$ amplitude. The $D^\ast\pi\{\tso\} \to D^\ast\pi\{\tdo\}$ amplitude is consistent with zero in the constrained energy region.

The importance of understanding $S$-wave interactions between pairs of hadrons in QCD cannot be understated. 
The strength of interactions observed here between a vector and a pseudoscalar not only extends our understanding of $D$-meson decays; it may point toward a resolution of some of the many other puzzles found with interacting charmed hadrons.


\acknowledgments
{
We thank our colleagues within the Hadron Spectrum Collaboration (www.hadspec.org).
DJW acknowledges support from a Royal Society University Research Fellowship. DJW acknowledges support from the U.K. Science and Technology Facilities Council (STFC) [grant number ST/T000694/1].
The software codes {\tt Chroma}~\cite{Edwards:2004sx}, {\tt QUDA}~\cite{Clark:2009wm,Babich:2010mu}, {\tt QPhiX}~\cite{Joo:2013lwm}, and {\tt QOPQDP}~\cite{Osborn:2010mb,Babich:2010qb} were used to compute the propagators required for this project.
This work used the Cambridge Service for Data Driven Discovery (CSD3), part of which is operated by the University of Cambridge Research Computing Service (www.csd3.cam.ac.uk) on behalf of the STFC DiRAC HPC Facility (www.dirac.ac.uk). The DiRAC component of CSD3 was funded by BEIS capital funding via STFC capital grants ST/P002307/1 and ST/R002452/1 and STFC operations grant ST/R00689X/1. Other components were provided by Dell EMC and Intel using Tier-2 funding from the Engineering and Physical Sciences Research Council (capital grant EP/P020259/1).
This work also used clusters at Jefferson Laboratory under the USQCD Initiative and the LQCD ARRA project, and the authors acknowledge support from the U.S. Department of Energy, Office of Science, Office of Advanced Scientific Computing Research and Office of Nuclear Physics, Scientific Discovery through Advanced Computing (SciDAC) program, and the U.S. Department of Energy Exascale Computing Project.
This research was supported in part under an ALCC award, and used resources of the Oak Ridge Leadership Computing Facility at the Oak Ridge National Laboratory, which is supported by the Office of Science of the U.S. Department of Energy under Contract No. DE-AC05-00OR22725. This research is also part of the Blue Waters sustained-petascale computing project, which is supported by the National Science Foundation (awards OCI-0725070 and ACI-1238993) and the state of Illinois. Blue Waters is a joint effort of the University of Illinois at Urbana-Champaign and its National Center for Supercomputing Applications. This work is also part of the PRAC “Lattice QCD on Blue Waters”. This research used resources of the National Energy Research Scientific Computing Center (NERSC), a DOE Office of Science User Facility supported by the Office of Science of the U.S. Department of Energy under Contract No. DEAC02-05CH11231. The authors acknowledge the Texas Advanced Computing Center (TACC) at The University of Texas at Austin for providing HPC resources that have contributed to the research results reported within this paper.
Gauge configurations were generated using resources awarded from the U.S. Department of Energy INCITE program at Oak Ridge National Lab, NERSC, the NSF Teragrid at the Texas Advanced Computer Center and the Pittsburgh Supercomputer Center, as well as at Jefferson Lab.
}


\bibliographystyle{apsrev4-1}
\bibliography{biblio.bib}


\clearpage

\begin{widetext}
\section*{Supplemental}

\subsection{Energy levels from other irreps}
\label{supp:spectra}

In Fig.~\ref{fig_spec_other} we present the remaining irreps used to constrain the amplitudes in addition to those presented in the main text.

\begin{figure*}[!thb]
    \includegraphics[width=1.0\textwidth]{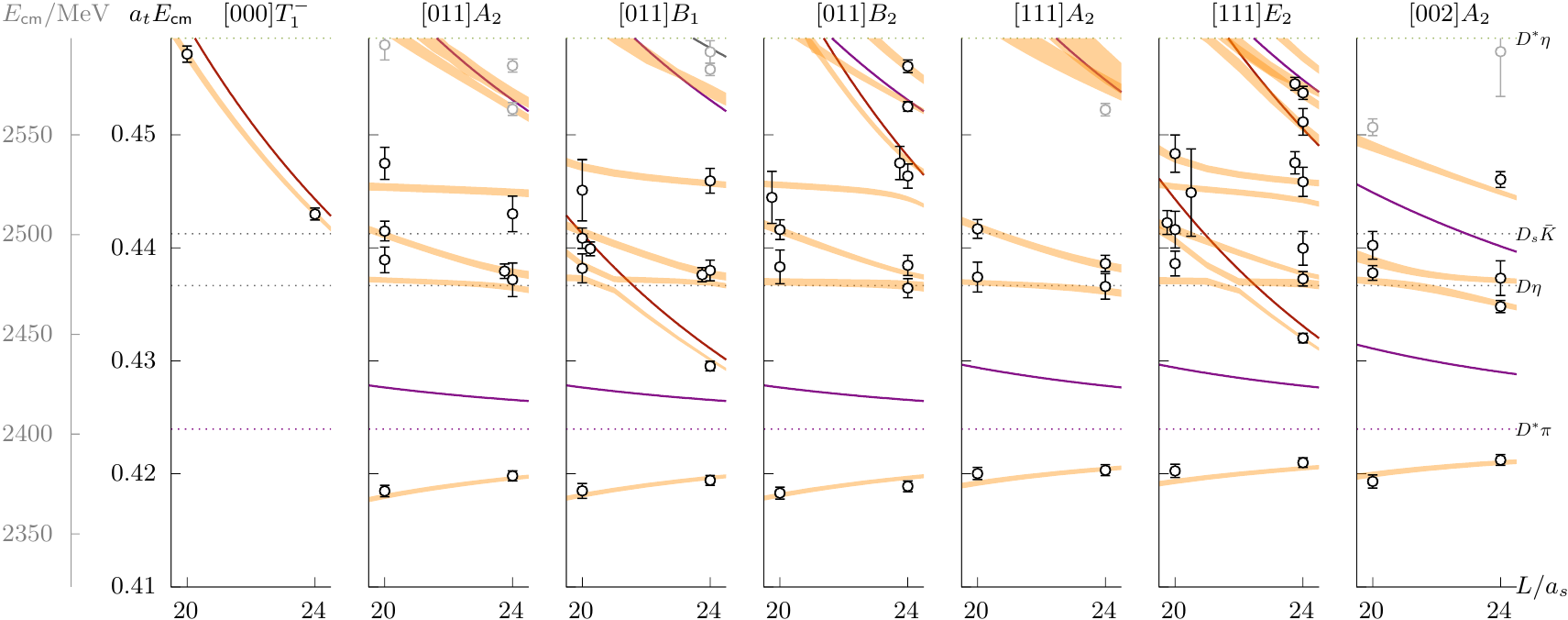}
	\caption{Energy levels and finite-volume solutions of the determinant condition from remaining irreps used in the fit, that are not shown in figure \ref{fig_spec}.}	
	\label{fig_spec_other}
\end{figure*}


\subsection{Operator tables}
\label{supp:ops}

In tables \ref{tab:ops1} and \ref{tab:ops2} we summarize the operators used to compute the correlation matrices from which the spectra are determined. Further details of the $\bar{q} \mathbf{\Gamma} q$ operators are given in Ref.~\cite{Dudek:2010wm}. The construction of two-body operators is described in Refs.~\cite{Thomas:2011rh,Dudek:2012gj}. Operators labeled $f_0$, $\rho$ and $D_0$ resemble three-body states as is described in Refs.~\cite{Woss:2019hse}. These are only found to contribute far above the energy region considered in this study.

\begin{table}[!htb]
	\begin{tabular}{lllllll}
		\toprule
		$[000] T_1^+$ & $[000]E^+$ & $[000]T_2^+$ & $[001]A_2$ & $[001] E_2$ & $[001] B_1$ & $[001] B_2$ \\
		\midrule
		$D_{[000]} \rho_{[000]}$ (1) & $D_{[100]} \pi_{[100]}$ (1) & $D_{[110]} \pi_{[110]}$ (1) & $D_{[100]} \rho_{[000]}$ (1) & $D_{[100]} \pi_{[110]}$ (1) & $D_{[100]} \pi_{[110]}$ (1) & $D_{[111]} \pi_{[110]}$ (1) \\
		$D_{[100]} \rho_{[100]}$ (2) & $D_{[110]} \pi_{[110]}$ (1) & ${D^*}_{[100]} \pi_{[100]}$ (1) & $D_{[000]} {f_0}_{[100]}$ (1) & $D_{[110]} \pi_{[100]}$ (1) & $D_{[110]} \pi_{[100]}$ (1) & $D_{[110]} {f_0}_{[100]}$ (1) \\
		$D_{[100]} {f_0}_{[100]}$ (1) & $D_{[200]} \pi_{[200]}$ (1) & ${D^*}_{[000]} \rho_{[000]}$ (1) & $D_{[100]} {f_0}_{[000]}$ (1) & $D_{[111]} \pi_{[110]}$ (1) & $D_{[100]} \eta_{[110]}$ (1) & ${D^*}_{[100]} \pi_{[110]}$ (2) \\
		${D^*}_{[000]} \pi_{[000]}$ (1) & $D_{[100]} \eta_{[100]}$ (1) & $\bar{q} \mathbf{\Gamma} q$ (29) & ${D^*}_{[000]} \pi_{[100]}$ (1) & $D_{[110]} \eta_{[100]}$ (1) & $D_{[110]} \eta_{[100]}$ (1) & ${D^*}_{[110]} \pi_{[100]}$ (2) \\
		${D^*}_{[100]} \pi_{[100]}$ (2) & $D_{[110]} \eta_{[110]}$ (1) &  & ${D^*}_{[100]} \pi_{[000]}$ (1) & $D_{[000]} \rho_{[100]}$ (1) & ${D_s}_{[100]} \bar{K}_{[110]}$ (1) & $\bar{q} \mathbf{\Gamma} q$ (20) \\
		${D^*}_{[110]} \pi_{[110]}$ (3) & ${D_s}_{[100]} \bar{K}_{[100]}$ (1) &  & ${D^*}_{[110]} \pi_{[100]}$ (2) & $D_{[100]} \rho_{[000]}$ (1) & ${D_s}_{[110]} \bar{K}_{[100]}$ (1) &  \\
		${D^*}_{[000]} \eta_{[000]}$ (1) & ${D_s}_{[110]} \bar{K}_{[110]}$ (1) &  & ${D^*}_{[100]} \eta_{[000]}$ (1) & ${D_s}_{[110]} \bar{K}_{[100]}$ (1) & $\bar{q} \mathbf{\Gamma} q$ (12) &  \\
		${D^*}_{[100]} \eta_{[100]}$ (2) & $\bar{q} \mathbf{\Gamma} q$ (4) &  & ${D_s^*}_{[100]} \bar{K}_{[000]}$ (1) & ${D^*}_{[000]} \pi_{[100]}$ (1) &  &  \\
		${D^*}_{[000]} \rho_{[000]}$ (1) &  &  & ${D_0}_{[100]} \pi_{[000]}$ (1) & ${D^*}_{[100]} \pi_{[000]}$ (1) &  &  \\
		${D_s^*}_{[000]} \bar{K}_{[000]}$ (1) &  &  & $\bar{q} \mathbf{\Gamma} q$ (32) & ${D^*}_{[110]} \pi_{[100]}$ (3) &  &  \\
		${D_s^*}_{[100]} \bar{K}_{[100]}$ (2) &  &  &  & ${D^*}_{[000]} \eta_{[100]}$ (1) &  &  \\
		${D_0}_{[100]} \pi_{[100]}$ (1) &  &  &  & ${D^*}_{[100]} \eta_{[000]}$ (1) &  &  \\
		$\bar{q} \mathbf{\Gamma} q$ (44) &  &  &  & ${D^*}_{[100]} {f_0}_{[000]}$ (1) &  &  \\
		&  &  &  & ${D_s^*}_{[100]} \bar{K}_{[000]}$ (1) &  &  \\
		&  &  &  & $\bar{q} \mathbf{\Gamma} q$ (44) &  &  \\
		\bottomrule
	\end{tabular}
	\caption{Operators included in the computation of the spectra shown in figure \ref{fig_spec}. For meson-meson operators, the numbers in parentheses indicate degeneracies in the non-interacting case. For the $\bar{q}q$-like operators (last line of each column), the number in parentheses gives the number of different $\Gamma$ constructions used.}
	\label{tab:ops1}
\end{table}

\begin{table}[!htb]
	\begin{tabular}{lllllll}
		\toprule
		$[000] T_1^-$ & $[011] A_2$ & $[011] B_1$ & $[011] B_2$ & $[111] A_2$ & $[111] E_2$ & $[002] A_2$ \\
		\midrule
		$D_{[100]} \pi_{[100]}$ (1) & $D_{[110]} \pi_{[110]}$ (1) & $D_{[100]} \pi_{[100]}$ (1) & $D_{[110]} \pi_{[110]}$ (1) & $D_{[111]} \rho_{[000]}$ (1) & $D_{[100]} \pi_{[110]}$ (1) & $D_{[100]} \rho_{[100]}$ (1) \\
		$D_{[100]} \eta_{[100]}$ (1) & $D_{[110]} \rho_{[000]}$ (1) & $D_{[110]} \pi_{[110]}$ (1) & $D_{[111]} \pi_{[100]}$ (1) & $D_{[111]} {f_0}_{[000]}$ (1) & $D_{[110]} \pi_{[100]}$ (1) & $D_{[100]} {f_0}_{[100]}$ (1) \\
		${D^*}_{[100]} \pi_{[100]}$ (1) & $D_{[110]} {f_0}_{[000]}$ (1) & $D_{[210]} \pi_{[100]}$ (1) & $D_{[110]} \rho_{[000]}$ (1) & ${D^*}_{[110]} \pi_{[100]}$ (2) & $D_{[211]} \pi_{[100]}$ (1) & $D_{[200]} {f_0}_{[000]}$ (1) \\
		$\bar{q} \mathbf{\Gamma} q$ (20) & ${D^*}_{[100]} \pi_{[100]}$ (2) & $D_{[100]} \eta_{[100]}$ (1) & $D_{[100]} {f_0}_{[100]}$ (1) & ${D^*}_{[111]} \pi_{[000]}$ (1) & $D_{[100]} \eta_{[110]}$ (1) & ${D^*}_{[100]} \pi_{[100]}$ (1) \\
		& ${D^*}_{[110]} \pi_{[000]}$ (1) & $D_{[110]} \rho_{[000]}$ (1) & ${D^*}_{[000]} \pi_{[110]}$ (1) & ${D^*}_{[111]} \eta_{[000]}$ (1) & $D_{[110]} \eta_{[100]}$ (1) & ${D^*}_{[200]} \pi_{[000]}$ (1) \\
		& ${D^*}_{[111]} \pi_{[100]}$ (2) & ${D_s}_{[100]} \bar{K}_{[100]}$ (1) & ${D^*}_{[100]} \pi_{[100]}$ (2) & ${D_s^*}_{[111]} \bar{K}_{[000]}$ (1) & $D_{[111]} \rho_{[000]}$ (1) & ${D^*}_{[210]} \pi_{[100]}$ (2) \\
		& ${D^*}_{[110]} \eta_{[000]}$ (1) & ${D^*}_{[000]} \pi_{[110]}$ (1) & ${D^*}_{[110]} \pi_{[000]}$ (1) & ${D_0}_{[111]} \pi_{[000]}$ (1) & $D_{[110]} {f_0}_{[100]}$ (1) & ${D^*}_{[100]} \eta_{[100]}$ (1) \\
		& ${D_s^*}_{[110]} \bar{K}_{[000]}$ (1) & ${D^*}_{[100]} \pi_{[100]}$ (1) & ${D^*}_{[111]} \pi_{[100]}$ (1) & $\bar{q} \mathbf{\Gamma} q$ (36) & ${D_s}_{[100]} \bar{K}_{[110]}$ (1) & ${D^*}_{[200]} \eta_{[000]}$ (1) \\
		& ${D_0}_{[110]} \pi_{[000]}$ (1) & ${D^*}_{[110]} \pi_{[000]}$ (1) & ${D^*}_{[110]} \eta_{[000]}$ (1) &  & ${D_s}_{[110]} \bar{K}_{[100]}$ (1) & ${D_s^*}_{[200]} \bar{K}_{[000]}$ (1) \\
		& $\bar{q} \mathbf{\Gamma} q$ (52) & ${D^*}_{[111]} \pi_{[100]}$ (2) & ${D_s^*}_{[110]} \bar{K}_{[000]}$ (1) &  & ${D^*}_{[100]} \pi_{[110]}$ (3) & $\bar{q} \mathbf{\Gamma} q$ (32) \\
		&  & ${D^*}_{[100]} \eta_{[100]}$ (1) & $\bar{q} \mathbf{\Gamma} q$ (52) &  & ${D^*}_{[110]} \pi_{[100]}$ (3) &  \\
		&  & ${D^*}_{[110]} \eta_{[000]}$ (1) &  &  & ${D^*}_{[111]} \pi_{[000]}$ (1) &  \\
		&  & ${D^*}_{[110]} {f_0}_{[000]}$ (1) &  &  & ${D^*}_{[111]} \eta_{[000]}$ (1) &  \\
		&  & ${D_s^*}_{[110]} \bar{K}_{[000]}$ (1) &  &  & ${D_s^*}_{[111]} \bar{K}_{[000]}$ (1) &  \\
		&  & $\bar{q} \mathbf{\Gamma} q$ (44) &  &  & $\bar{q} \mathbf{\Gamma} q$ (60) &  \\
		\bottomrule
	\end{tabular}
	\caption{Same as table \ref{tab:ops1} but for the spectra shown in figure \ref{fig_spec_other}. }
	\label{tab:ops2}
\end{table}


\FloatBarrier
\subsection{Reference parameterization}
\label{sec:ref_param}

The reference parameterization consists of partial wave amplitudes in $J^P=1^+$, $2^+$, $0^-$, $1^-$ and $2^-$. Both $D\pi$ and $D^\ast\pi$ contribute in various combinations of total spin $J$, angular momentum $\ell$ and spin $S$, identified below as $^{2S+1}\ell_J$ (spectroscopic notation). Although the focus of this study is \emph{elastic} $D^\ast\pi$ scattering, each partial wave $J^P$ can receive contributions from multiple $^{2S+1}\ell_J$ combinations which can couple. For $J^P=1^+$, we have
\begin{align}
\bm{t} {(J^P=1^+)}=\left(
\begin{matrix}
t(D^\ast\pi\{\tso\}\to D^\ast\pi\{\tso\}) & t(D^\ast\pi\{\tso\}\to D^\ast\pi\{\tdo\}) \\
t(D^\ast\pi\{\tdo\}\to D^\ast\pi\{\tso\}) & t(D^\ast\pi\{\tdo\}\to D^\ast\pi\{\tdo\}) 
\end{matrix}
\right)\,.
\end{align}
For $J^P=2^+$, only one $D^\ast\pi$ combination arises, however $D\pi$ also contributes,
\begin{align}
\bm{t} {(J^P=2^+)}=\left(
\begin{matrix}
t(D\pi\{\odt\}\to D\pi\{\odt\})      & t(D\pi\{\odt\}\to D^\ast\pi\{\tdt\}) \\
t(D^\ast\pi\{\tdt\}\to D\pi\{\odt\}) & t(D^\ast\pi\{\tdt\}\to D^\ast\pi\{\tdt\}) 
\end{matrix}
\right)\,.
\end{align}
For $J^P=0^-$ and $2^-$, we consider just a single $t(D^\ast\pi\{^3P_J\}\to D^\ast\pi\{^3P_J\})$ combination. In $J^P=1^-$, there are two possible $\ell=1$ combinations,
\begin{align}
\bm{t} {(J^P=1^-)}=\left(
\begin{matrix}
t(D\pi\{\opo\}\to D\pi\{\opo\})      & t(D\pi\{\opo\}\to D^\ast\pi\{\tpo\}) \\
t(D^\ast\pi\{\tpo\}\to D\pi\{\opo\}) & t(D^\ast\pi\{\tpo\}\to D^\ast\pi\{\tpo\}) 
\end{matrix}
\right)\,.
\end{align}

We parameterize each $J^P$ using a $K$-matrix coupling the relevant hadron-hadron $^{2S+1}\ell_J$ combinations. For the $J^P=1^+$ partial wave, we use
\begin{align}
K_{ij}=\frac{g_{0,i}g_{0,j}}{m_0^2-s}+\frac{g_{1,i}g_{1,j}}{m_1^2-s}+\gamma_{ij}
\label{eq:param_1p}
\end{align}
where $i,j$ refer to the $^{2S+1}\ell_J$ hadron-hadron combinations. In this study $^3\ell_J$ is always $D^\ast\pi$ and $^1\ell_J$ is always $D\pi$. Any parameter not given a value below, such as many of the $\gamma_{ij}$, is implicitly fixed to zero. Where we specify the channel label $i$ explicitly using spectroscopic notation, we write $g_p(i)$ instead of $g_{p,i}$ and $\gamma(i\to j)$ instead of $\gamma_{ij}$. Superscripts in parentheses are used for coefficients of $s$ and indicate the order of the term.

These $K$-matrix parameters are inserted into a $t$-matrix that includes momenta raised to the power required by the near-threshold behaviour, $k_i^\ell$, appropriate for each element,
\begin{align}
t_{ij}^{-1}=\frac{1}{(2k_i)^{\ell_i}}K^{-1}_{ij}\frac{1}{(2k_j)^{\ell_j}}+I_{ij}
\label{eq:tmat_cm}
\end{align}
where $I_{ij}$ is a Chew-Mandelstam phase space as described in appendix B of Ref.~\cite{Wilson:2014cna}.

The $K$-matrices for the remaining amplitudes are similar. For $J^P=2^+$ we use just a single pole,
\begin{align}
K_{ij}=\frac{g_{2,i}g_{2,j}}{m_2^2-s}+\gamma_{ij}
\label{eq:param_2p}
\end{align}
where the channels $i,j$ are $D\pi(^1D_2)$ and $D^\ast\pi(^3D_2)$.

The negative parity waves are parameterized by constants $K_{ij}=\gamma_{ij}$. The corresponding plots of $\rho_i \rho_j |t_{ij}|^2$ can be seen in figure \ref{fig_amps2}. The $0^-$ and $2^-$ waves receive contributions from only $D^\ast\pi$. The $1^-$ wave has contributions from both $D\pi\{^1P_1\}$ and $D^\ast\pi\{^3P_1\}$. The $1^-$ amplitude contains the $D^\ast$ bound state far below threshold. We exclude this level from all the amplitude determinations, as there is not a large obvious effect on the spectrum above threshold. However, the parameterization is capable of describing an effect in the physically-allowed $D\pi$ and $D^\ast\pi$ scattering region above threshold. Further details of the $D^\ast$ bound-state on these lattices can be found in Ref.~\cite{Moir:2016srx}.

\vspace{1mm}
The parameter values and their correlations are:
{
\setcounter{MaxMatrixCols}{15}
\footnotesize
\begin{center}
\begin{tabular}{rll}
$m_0=$ & $(0.42294 \pm 0.00017 \pm 0.00011)\,\cdot\,a_t^{-1}$ & \multirow{7}{*}{ $\left.\rule{0cm}{1.3cm}\right\}$ $J^P=1^+$}\\ 
$m_1=$ & $(0.43691 \pm 0.00028 \pm 0.00006)\,\cdot\,a_t^{-1}$ & \\
$g_0({\tso})=$ & $(0.528 \pm 0.022 \pm 0.022)\,\cdot\,a_t^{-1}$ & \\
$g_0({\tdo})=$ & $(0.390 \pm 0.850 \pm 0.089)\,\cdot\,a_t$ & \\
$g_1({\tso})=$ & $(0.014 \pm 0.012 \pm 0.002)\,\cdot\,a_t^{-1}$ & \\
$g_1({\tdo})=$ & $(-6.40 \pm 1.36  \pm 0.21)\,\cdot\,a_t$ & \\
$\gamma(\tso\to\tso)=$ & $(10.3 \pm 1.2 \pm 0.6)$ & \\
$m_2=$ & $(0.44546 \pm 0.00029 \pm 0.00006)\,\cdot\,a_t^{-1}$ & \multirow{4}{*}{ $\left.\rule{0cm}{0.7cm}\right\}$ $J^P=2^+$}\\ 
$g_2(\odt)=$ & $(1.730 \pm 0.065 \pm 0.013)\,\cdot\,a_t$ & \\
$g_2(\tdt)=$ & $(3.02 \pm 0.92 \pm 0.09)\,\cdot\,a_t$ & \\
$\gamma(\odt\to\odt)=$ & $(222 \pm 172 \pm 225)\,\cdot\,a_t^4$ & \\
$\gamma(\tpz\to\tpz)=$  & $(-62 \pm 54 \pm 24)\,\cdot\,a_t^2$ & $\;\:\left.\right\}$ $J^P=0^-$\\
$\gamma(\opo\to\opo)=$  & $(15.3 \pm 3.3 \pm 10.2)\,\cdot\,a_t^2$ & \multirow{2}{*}{ $\left.\rule{0cm}{0.3cm}\right\}$ $J^P=1^-$}\\
$\gamma(\tpo\to\tpo)=$ & $(-107 \pm 16 \pm 18)\,\cdot\,a_t^2$ & \\
$\gamma(\tpt\to\tpt)=$ & $(86 \pm 21 \pm 18)\,\cdot\,a_t^2$ & $\;\:\left.\right\}$ $J^P=2^-$\\[1.3ex]
\end{tabular}
\end{center}

\begin{center}
\begin{tabular}{l}
$\begin{bmatrix*}[r]   1.00 &   0.38 &  \tm0.22 &  \tm0.05 &  \tm0.10 &  \tm0.00 &  \tm0.16 &   0.28 &  \tm0.02 &  \tm0.04 &  \tm0.13 &  \tm0.10 &  \tm0.32 &  \tm0.15 &  \tm0.12\\
&  1.00 &   0.01 &  \tm0.02 &   0.16 &   0.32 &  \tm0.02 &   0.34 &   0.11 &  \tm0.17 &  \tm0.11 &   0.12 &  \tm0.23 &  \tm0.04 &  \tm0.31\\
&&  1.00 &  \tm0.44 &  \tm0.03 &   0.40 &   0.96 &   0.03 &   0.03 &   0.28 &   0.18 &   0.36 &   0.10 &  \tm0.24 &  \tm0.24\\
&&&  1.00 &   0.52 &  \tm0.61 &  \tm0.54 &  \tm0.01 &   0.01 &  \tm0.34 &  \tm0.05 &  \tm0.06 &   0.01 &   0.41 &   0.26\\
&&&&  1.00 &  \tm0.06 &  \tm0.05 &  \tm0.03 &  \tm0.02 &  \tm0.07 &   0.08 &   0.27 &   0.03 &   0.12 &   0.01\\
&&&&&  1.00 &   0.40 &   0.01 &   0.08 &  \tm0.03 &  \tm0.05 &   0.39 &   0.01 &  \tm0.11 &  \tm0.71\\
&&&&&&  1.00 &  \tm0.00 &  \tm0.00 &   0.39 &   0.23 &   0.32 &   0.11 &  \tm0.32 &  \tm0.21\\
&&&&&&&  1.00 &   0.11 &  \tm0.03 &   0.00 &  \tm0.02 &  \tm0.18 &  \tm0.09 &  \tm0.10\\
&&&&&&&&  1.00 &  \tm0.17 &  \tm0.17 &   0.04 &   0.08 &  \tm0.03 &  \tm0.10\\
&&&&&&&&&  1.00 &   0.39 &  \tm0.16 &   0.01 &  \tm0.19 &   0.40\\
&&&&&&&&&&  1.00 &   0.00 &  \tm0.16 &  \tm0.06 &   0.24\\
&&&&&&&&&&&  1.00 &   0.06 &   0.04 &  \tm0.48\\
&&&&&&&&&&&&  1.00 &   0.14 &   0.11\\
&&&&&&&&&&&&&  1.00 &  \tm0.04\\
&&&&&&&&&&&&&&  1.00\end{bmatrix*}$  \\
\end{tabular}
\end{center}
}
\begin{align}
\chi^2/ N_\mathrm{dof} = \frac{95.0}{94-15} = 1.20\,.
\end{align}

The first uncertainty on the parameter values is the from the $\chi^2$ minimum. The second uncertainty is the largest deviation found between the central values of the parameters after varying the scattering hadron masses and anisotropy within their uncertainties.

\begin{figure}[!t]
	\includegraphics[width=1.0\textwidth]{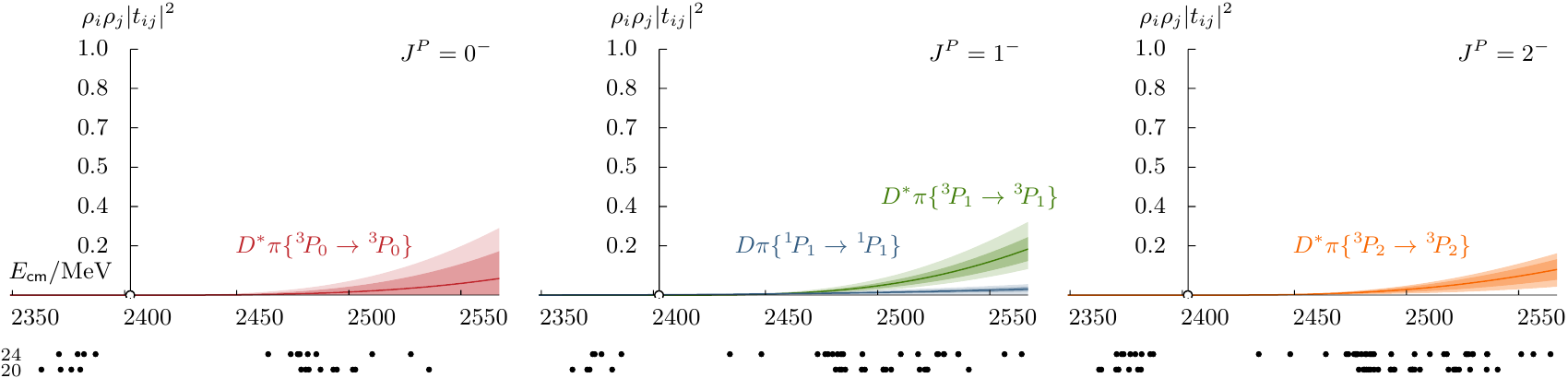}
	\caption{As figure \ref{fig_amps} but for $J^P \in \{0^-$, $1^-$, $2^-\}$.}	
	\label{fig_amps2}
\end{figure}


\subsection{Parameterization variations}
\label{supp:par_var}

In addition to the parameterization defined in section \ref{sec:ref_param}, we use the following amplitude variations. These include terms linear in $s$ and pole terms with numerators linear in $s$. All parameterizations use the $K$-matrix formalism and each block of the $K$-matrix corresponding to a $J^P$ combination can be given by one master formula, in which various parameters may be fixed to zero or floated in the fit. The complete list is given in table \ref{tab:params}.
For the $J^P = 1^+$ block of the $K$-matrix we use
\begin{equation}
	K_{ij} = \sum_{p \in \{0,1\}} \frac{\big(g_{p,i} + g^{(1)}_{p,i} s\big) \big(g_{p,j} + g^{(1)}_{p,j} s\big)}{m_p^2-s} + \gamma_{ij}  + \gamma^{(1)}_{ij} s \;.
	\label{eq:param_1p_ling}
\end{equation}
Here $i$, $j$ represent the two mixing partial waves of $D^\ast \pi$, $\{\tso\}$ and $\{\tdo\}$.
For the $J^P = 2^+$ block we have
\begin{equation}
	K_{ij}=\frac{\big(g_{2,i} + g^{(1)}_{2,i} s\big) \big(g_{2,j} + g^{(1)}_{2,j} s\big)}{m_2^2-s}  +\gamma_{ij}  + \gamma^{(1)}_{ij} s \;.
	\label{eq:param_2p_ling}
\end{equation}
The two channels represented by $i$, $j$ are $D\pi \{\odt\}$ and $D^\ast \pi \{\tdt\}$. The masses $m_0$, $m_1$ and $m_2$ in all $J^P=1^+$ and $J^P=2^+$ parameterizations are always free fit parameters.
For the $J^P = 0^-$ amplitude, corresponding to $D^\ast \pi \{\tpz \rightarrow \tpz\}$, we attempt fits of both a $K$-matrix including a pole term and one with a constant term only. The general $K$-matrix reads
\begin{align}
	K =\frac{g^2_3(\tpz)}{m_3^2-s}+\gamma(\tpz \rightarrow \tpz) \; .
	\label{eq:param_0m}
\end{align}
When the pole term is included, i.e. $g^{(3)}(\tpz) \neq 0$, the mass is fixed to $m_3 = 0.4707 \cdot a_t^{-1}$, based on observations from the $[000] A_1^-$ irrep as shown in Ref.~\cite{Moir:2013ub}. There is not sufficient constraint to allow $m_3$ to float.
The $J^P=1^-$ and $J^P=2^-$ amplitudes are well-described by a $K$-matrix with constant and linear terms only. For $J^P=1^-$ we have
\begin{equation}
	K_{ij}=  \gamma_{ij}  \; ,
	\label{eq:param_1m}
\end{equation}
where $i$, $j$ represent combinations of $D \pi \{\opo\}$ and $D^\ast \pi \{\tpo\}$. For $J^P=2^-$ we have a single channel only and the $K$-matrix reads
\begin{equation}
	K =  \gamma(\tpt \rightarrow \tpt)  + \gamma^{(1)}(\tpt \rightarrow \tpt) s \;.
	\label{eq:param_2m}
\end{equation}
Most parameterizations use the Chew-Mandelstam prescription (indicated by CM in table \ref{tab:params}), as given in equation \ref{eq:tmat_cm}. For the subtraction point we typically choose the mass parameter of the lowest pole in the respective channel. Further details are given in appendix B of Ref.~\cite{Wilson:2014cna}. For some amplitudes, we subtract at threshold instead. We also fit amplitude variations that use a simple phase space $\rho_i = 2k_i/\sqrt{s}$, with a $t$-matrix given by
\begin{align}
	t_{ij}^{-1}=\frac{1}{(2k_i)^{\ell_i}}K^{-1}_{ij}\frac{1}{(2k_j)^{\ell_j}} - i \rho_{i} \delta_{ij} \;.
	\label{eq:tmat_nocm}
\end{align}

We only include amplitude determinations in table \ref{tab:params} which have an acceptable fit quality and correspond to amplitudes with have no unphysical features such as complex poles on the first (physical) Riemann sheet.

\begin{table}[!htb]
	
	\begin{tabular}{lclrr}
		\toprule
		$J^P$ & Parameterization & Free parameters (couplings and polynomial) & $N_{\text{pars}}$ & $\chi^2/N_{DoF}$ \\
		\midrule
		\vspace{0.0cm} & & & &\\
		\multirow{11}{*}{$1^+$} &
		\multirow{11}{*}{\parbox{0.15\textwidth}{eq. \ref{eq:param_1p_ling} \\ (CM, pole 0) \hspace{0.2cm}}}  &  $\{ g_0 (\tdo),\;g_1 (\tdo),\;g_0 (\tso),\;g_1 (\tso),\;\gamma(\tso\to\tso) \}$ &                15 &             \bf{1.20} \\
		&                 &                              $\{ g_0^{(1)} (\tdo),\;g_1 (\tdo),\;g_0 (\tso),\;g_1 (\tso),\;\gamma(\tso\to\tso) \}$ &                15 &             1.20 \\
		&                 &                                                 $\{ g_1 (\tdo),\;g_0 (\tso),\;g_1 (\tso),\;\gamma(\tso\to\tso) \}$ &                14 &             1.19 \\
		&                 &                              $\{ g_0 (\tdo),\;g_1^{(1)} (\tdo),\;g_0 (\tso),\;g_1 (\tso),\;\gamma(\tso\to\tso) \}$ &                15 &             1.21 \\
		&                 &                                                 $\{ g_0 (\tdo),\;g_0 (\tso),\;g_1 (\tso),\;\gamma(\tso\to\tso) \}$ &                14 &              \emph{1.24} \\
		&                 &                              $\{ g_0 (\tdo),\;g_1 (\tdo),\;g_0^{(1)} (\tso),\;g_1 (\tso),\;\gamma(\tso\to\tso) \}$ &                15 &             1.20 \\
		&                 &                                                 $\{ g_0 (\tdo),\;g_1 (\tdo),\;g_0 (\tso),\;\gamma(\tso\to\tso) \}$ &                14 &             1.20 \\
		&                 &                 $\{ g_0 (\tdo),\;g_1 (\tdo),\;g_0 (\tso),\;g_0^{(1)} (\tso),\;g_1 (\tso),\;\gamma(\tso\to\tso) \}$ &                16 &             1.22 \\
		&                 &  $\{ g_0 (\tdo),\;g_1 (\tdo),\;g_0 (\tso),\;g_1 (\tso),\;\gamma(\tso\leftrightarrow\tdo),\;\gamma(\tso\to\tso) \}$ &                16 &             1.16 \\
		&                 &        $\{ g_0 (\tdo),\;g_1 (\tdo),\;g_0 (\tso),\;g_1 (\tso),\;\gamma(\tso\to\tso),\;\gamma^{(1)}(\tso\to\tso) \}$ &                16 &             1.21 \\
		&                 &                                                          $\{ g_0 (\tdo),\;g_1 (\tdo),\;g_0 (\tso),\;g_1 (\tso) \}$ &                14 &             \emph{1.35} \\
		
		\vspace{0.1cm} & & & &\\
		
		\midrule
		\vspace{0.0cm} & & & &\\
		\multirow{10}{*}{$2^+$} &
		\multirow{5}{*}{\parbox{0.15\textwidth}{eq. \ref{eq:param_2p_ling}\\ (CM, pole 2)  \hspace{0.2cm}}} &              $\{ g_2 (\odt),\;g_2 (\tdt),\;\gamma(\odt\to\odt) \}$ &                15 &             \bf{1.20} \\
		&                 &  $\{ g_2^{(1)} (\odt),\;g_2 (\tdt),\;\gamma(\odt\to\odt) \}$ &                15 &             1.20 \\
		&                 &  $\{ g_2 (\odt),\;g_2^{(1)} (\tdt),\;\gamma(\odt\to\odt) \}$ &                15 &             1.20 \\
		&                 &                              $\{ g_2 (\odt),\;g_2 (\tdt) \}$ &                14 &             1.21 \\
		&                 &  $\{ g_2 (\odt),\;g_2 (\tdt),\;\gamma^{(1)}(\odt\to\odt) \}$ &                15 &             1.20 \\
		\vspace{0.0cm} & & & &\\
		& \parbox{0.15\textwidth}{(CM, thresh.)} &  $\{ g_2 (\odt),\;g_2 (\tdt),\;\gamma(\odt\to\odt) \}$ &                15 &             1.20 \\
		\vspace{0.0cm} & & & &\\
		& \parbox{0.15\textwidth}{(w/o CM)} & $\{ g_2 (\odt),\;g_2 (\tdt),\;\gamma(\odt\to\odt) \}$ &                15 &             1.20 \\
		\vspace{0.1cm} & & & &\\
		
		\midrule
		\vspace{0.0cm} & & & &\\
		\multirow{5}{*}{$0^-$} &
		\parbox{0.15\textwidth}{eq. \ref{eq:param_0m} \\ (CM, thresh.)  \hspace{0.2cm}} &  $\{ \gamma(\tpz \to \tpz) \}$ &                15 &             \bf{1.20} \\
		\vspace{0.0cm} & & & &\\
		& \parbox{0.15\textwidth}{(CM, pole 3)  \hspace{0.2cm}} &  $\{ g_3(\tpz),\; \gamma(\tpz \to \tpz) \}$ &                16 &             1.22 \\
		\vspace{0.0cm} & & & &\\
		& \parbox{0.15\textwidth}{no $J^P =0^-$  \hspace{0.2cm}} &  - &                14 &             1.20 \\
		\vspace{0.1cm} & & & &\\
		
		\midrule
		\vspace{0.0cm} & & & &\\
		\multirow{4}{*}{$1^-$} &
		\multirow{2}{*}{\parbox{0.15\textwidth}{eq. \ref{eq:param_1m} \\ (CM, thresh.)  \hspace{0.2cm}}} &  $\{ \gamma(\opo \to \opo),\; \gamma(\tpo \to \tpo) \}$ &                15 &             \bf{1.20} \\
		& &  $\{ \gamma(\opo \to \opo),\; \gamma(\tpo \to \tpo),\; \gamma(\opo \leftrightarrow \tpo) \}$ &                16 &             1.20 \\
		\vspace{0.0cm} & & & &\\
		& \parbox{0.15\textwidth}{(w/o CM)  \hspace{0.2cm}} &  $\{ \gamma(\opo \to \opo),\; \gamma(\tpo \to \tpo) \}$ &                15 &             1.20 \\
		\vspace{0.1cm} & & & &\\
		
		\midrule
		\vspace{0.0cm} & & & &\\
		\multirow{3}{*}{$2^-$} &
		\multirow{3}{*}{\parbox{0.15\textwidth}{eq. \ref{eq:param_2m} \\ (CM, thresh.)  \hspace{0.2cm}}} &  $\{ \gamma(\tpt \to \tpt) \}$ &                15 &             \bf{1.20} \\
		& &  $\{ \gamma(\tpt \to \tpt),\; \gamma^{(1)}(\tpt \to \tpt) \}$ &                16 &             1.19 \\
		& &  $\{ \gamma^{(1)}(\tpt \to \tpt) \}$  &                15 &             1.21 \\
		
		\vspace{0.2cm} & & & &\\
		\bottomrule
	\end{tabular}
	\caption{Amplitude variations and the corresponding fit results. The first row of each block belonging to one combination of $J^P$ corresponds to the reference parameterization described in section \ref{sec:ref_param} (last column printed in bold face). The free parameters belong to the respective referenced equation. The phase space prescription is given in parentheses. For each variation in a given $J^P$ block of the $K$-matrix, all other blocks correspond to the reference parameterization. Couplings and polynomial coefficients not listed under ``Free parameters'' are fixed to zero. Amplitudes, where the entry in the last column is printed in italics, were removed from the calculation of the final result due to a comparatively large $\chi^2$-minimum (see also section \ref{sec:discarded_param}).}
	\label{tab:params}
	
\end{table}

\end{widetext}
\FloatBarrier


\subsection{Parameterization with $g_1(\tdo)=0$}
\label{sec:discarded_param}

\begin{figure}[th]
    \vspace{0.3cm}
    \includegraphics[width=0.97\columnwidth]{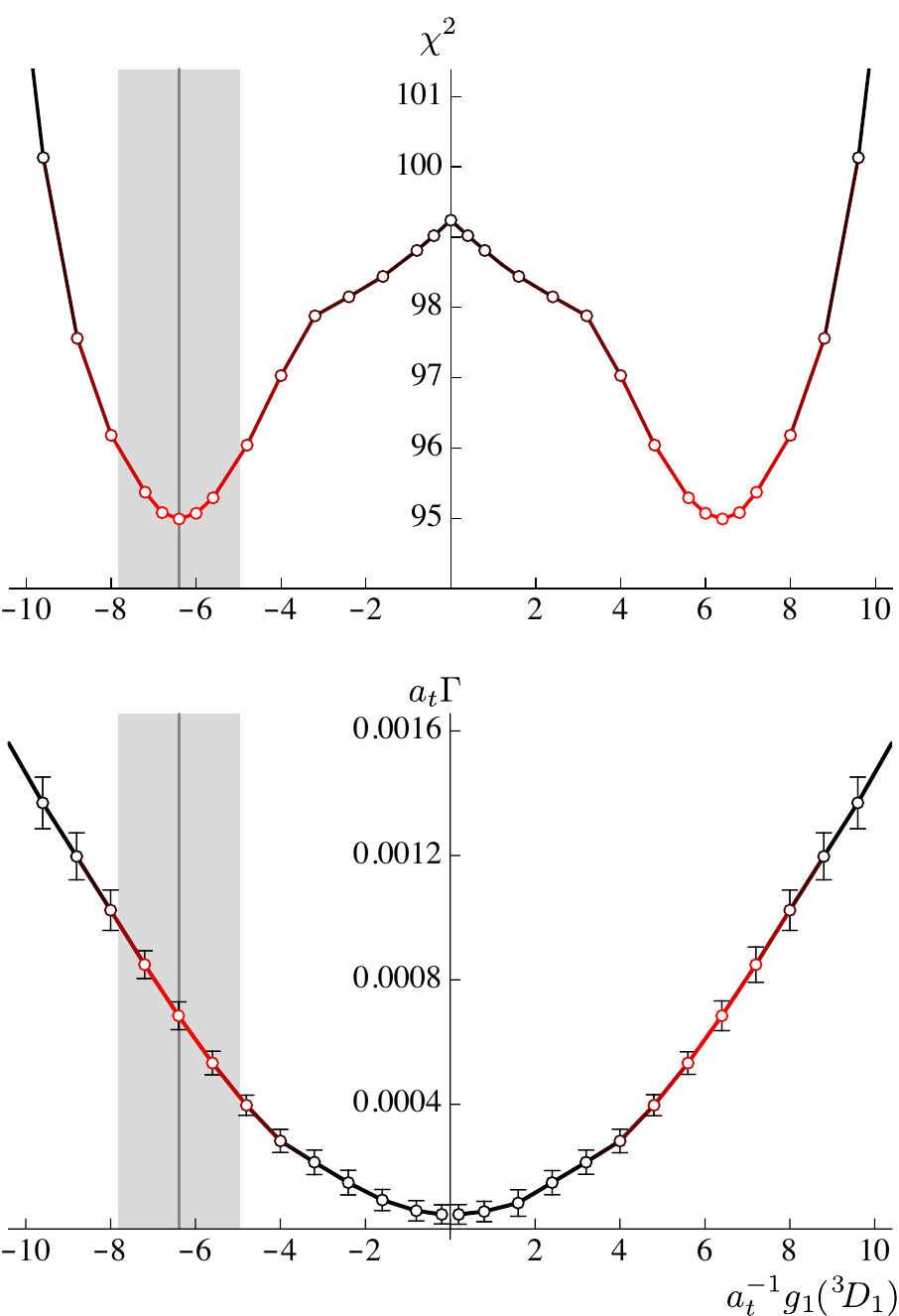}
    \vspace{0.3cm}
	\caption{Upper panel: The $\chi^2$ value obtained after minimization as a function of $g_1({\tdo})$. The vertical grey band shows the value and uncertainty on this parameter obtained from the determination of the reference parameterization, $g_1({\tdo})= (-6.40 \pm 1.36) \cdot a_t$. Lower panel: The width of the pole $\Gamma=-2\;\mathrm{Im}\sqrt{s_0}$ as a function of $g_1({\tdo})$. The open circular markers indicate values of $g_1({\tdo})$ where a fit was performed, the solid red-black line in both panels is for visual guidance only.}	
	\label{fig_chisq_v_g}
\end{figure}

In one of the parameterization variations, the $K$-matrix pole-coupling parameter to the $\tdo$ amplitude for the pole with a higher mass is fixed to $g_1(\tdo)=0$. This differs from the reference parameterization in section~\ref{sec:ref_param} only by the change to this parameter. After minimization, we find $\chi^2/N_\mathrm{dof}=\frac{99.24}{94 - 14} = 1.24$, slightly larger than the reference parameterization but not unreasonably large. The imaginary part of the higher $J^P=1^+$ resonance pole found in this amplitude around $a_t\mathrm{Re}\sqrt{s_0}\approx 0.437$ is much smaller compared to the other parameterizations, which requires further investigation.

This parameterization and the reference parameterization can be interpolated by adjusting $g_1(\tdo)$ from the value in the reference parameterization $g_1({\tdo})= (-6.40 \pm 1.36) \cdot a_t$ to zero. We adjust $g_1({\tdo})$ in small steps and determine the remaining parameters using the $\chi^2$ minimization procedure.

In Fig.~\ref{fig_chisq_v_g}, the $\chi^2$ value and $\Gamma=-2\:\mathrm{Im}\sqrt{s_0}$ are shown as a function of $g_1({\tdo})$. As expected, we see a clear minimum in the $\chi^2$ at the value of $g_1({\tdo})$ obtained from the reference parameterization. There is a symmetry in $g_1({\tdo})\to -g_1({\tdo})$. $K_{ij}$ is invariant under certain sign changes, in particular when both $g_1({\tso})$ and $g_1({\tdo})$ change sign simultaneously. The relative sign of $g_1({\tso})$ and $g_1({\tdo})$ is relevant, as are the signs of the $\gamma$ parameters. The imaginary part of the pole position determined from the minima is approximately quadratic in $g_1({\tdo})$. 

We see that the spectra clearly favour a non-zero value, and that $g_1({\tdo})=0$ is not special. However, at the level of a few standard deviations, we cannot exclude a vanishingly small value. This is reflected in the quoted value of the width in the main text, $\Gamma=(5\pm3)$~MeV. 

A better determination of the width of this pole could be obtained by utilizing larger volumes. This would provide additional non-interacting meson-meson energies in the region of this state. The avoided level crossings induced by the resonance would then improve the amplitude determination.


\end{document}